\documentclass[prl,amsmath,amssymb,superscriptaddress,twocolumn]{revtex4}
\usepackage{graphicx}
\usepackage{bbm}

\newcommand{\br}{{\bf r}}
\newcommand{\bx}{{\bf x}}
\newcommand{\by}{{\bf y}}
\newcommand{\bz}{{\bf z}}
\newcommand{\bk}{{\bf k}}

\newcommand{\bp}{{\bf p}}
\newcommand{\bv}{{\bf v}}

\newcommand{\bA}{{\bf A}}

\newcommand{\bG}{{\bf G}}

\newcommand{\bdelta}{\delta}

\begin{document}
\title{Index theoretic characterization of $d$-wave superconductors in the vortex state}
\author{Oskar Vafek}
\affiliation{Stanford University Institute for Theoretical Physics
and Department of Physics, Stanford, CA 94305}
\author{Ashot Melikyan}
\affiliation{Institute of Fundamental Theory, Department of Physics,
University of Florida, Gainesville, FL 32611}
\date{\today}
\begin{abstract}
We employ index theoretic methods to study analytically the low
energy spectrum of a lattice $d$-wave superconductor in the vortex
lattice state. This allows us to compare singly quantized $hc/2e$
and doubly quantized $hc/e$ vortices, the first of which must always
be accompanied by $Z_2$ branch cuts. For an inversion symmetric
vortex lattice and in the presence of particle-hole symmetry we
prove an index theorem that imposes a lower bound on the number of
zero energy modes. Generic cases are constructed in which this bound
exceeds the number of zero modes of an equivalent lattice of doubly
quantized vortices, despite the identical point group symmetries.
The quasiparticle spectrum around the zero modes is doubly
degenerate and exhibits a Dirac-like dispersion, with velocities
that become universal functions of $\Delta_0/t$ in the limit of low
magnetic field. For weak particle-hole symmetry breaking, the gapped
state can be characterized by a topological quantum number, related
to spin Hall conductivity, which generally differs in the cases of
the $hc/2e$ and $hc/e$ vortex lattices.
\end{abstract} \maketitle

Understanding the interactions among fermionic quasiparticles and
topological excitations of two dimensional superconductors
constitutes one of the main challenges in modern condensed matter
physics. It is almost certainly true that at least some of the
peculiar phenomena observed in the underdoped cuprates are related
to the existence of nodal fermionic excitations of a
$d_{x^2-y^2}$wave superconductor with low superfluid density, which
is eventually susceptible to dephasing by proliferation of mobile
vortices \cite{EmeryKivelson1995}.

When an external magnetic field induces a finite density of vortices
which form an Abrikosov lattice, the motion of vortices can be
neglected and the problem of the quasiparticle spectrum is
simplified. The initial theoretical analysis was based on numerical
computations \cite{WangMacDonald1995}, semiclassical approximation
\cite{Volovik1993}, and scaling arguments \cite{SimonLee1997}. As
pointed out by Franz and
Tesanovic\cite{FranzTesanovic2000,VafekMelikyan1_2001,VafekMelikyan2_2001},
the quasiparticles interact with the $hc/2e$ vortices by {\em both}
the (semiclassical) coupling of the charge currents to the superflow
as well as a purely quantum coupling of the spin currents to the
$Z_2$ branch cuts emanating from the vortices. Marinelli, Halperin,
and Simon \cite{Marinelli2000} then showed that within the
linearized approximation the spectrum at each node remains gapless
to all orders of perturbation theory in phase gradients. This result
was rederived by an insightful symmetry analysis in Ref.
\cite{Ashvin2001}, where it was concluded that the effect of the
$hc/2e$ vortices is merely a possible renormalization of the nodal
quasiparticle velocities, but that the presence and particularly
{\em the number of} the zero modes is the same as in the lattice of
doubly quantized $hc/e$ vortices. Various topological aspects of the
problem were discussed in
\cite{Morita2001,Ashvin2001,VafekMelikyan2_2001}.

In this Letter, we revisit the problem of the Fermionic
(quasiparticle) spectrum of a lattice $d$-wave superconductor in the
vortex state. Our main result is that there can arise a fundamental
topological difference between the singly and doubly quantized
vortex lattice which is beyond perturbative analysis
\cite{Marinelli2000,Ashvin2001}. In the case of an inversion
symmetric vortex lattice, this difference relates to the non-trivial
transformation properties of the $Z_2$ branchcuts under inversion,
absent for $hc/e$ vortices.

Our point of departure is the inversion symmetric vortex lattice
state of the $d$-wave superconductor on a tight-binding lattice with
perfect particle-hole (p-h) symmetry. We use the p-h symmetry to
conduct non-perturbative analysis of the low energy spectrum and
find conditions sufficient for the existence of zero energy nodal
states. Armed with this analysis, we study weak breaking of the p-h
symmetry, which gaps the nodal points, and show that the resulting
state is topologically non-trivial. It is characterized by a
topological quantum number proportional to the spin-Hall
conductivity
\cite{Volovik89,Senthil1999,Ashvin2001,VafekMelikyan2_2001}.

% equal to $\pm4$ for $hc/2e$ lattice, as opposed to $\pm2$ for
%$hc/e$ lattice, the latter being topologically equivalent to $d\pm
%id$ \cite{Laughlin98}, the former to $g\pm ig$, superconductor
%\cite{gig}.

Remarkably, in cases when the branch cuts transform non-trivially
under inversion, the number of the zero modes is doubled for $hc/2e$
vortices relative to the $hc/e$ counterpart, despite identical
discrete translational symmetry of both vortex lattices in question.
The difference also holds for a lattice of $hc/2e$ vortices, when
the branch cuts are ignored. We demonstrate this by proving a form
of an index theorem \cite{AtiyahSinger68,WenZee89}, which puts a
lower bound on the number of zero modes, and by explicit analytical
and numerical solutions, which reveal that the lower bound is, in
fact, saturated.

We now provide justification for the above claims. Consider the
Hamiltonian $H=H_0-\mu N$, where
\begin{equation}\label{h0}
H_0=\sum_{\langle\br\br'\rangle}\left[
-t_{\br\br'}c_{\br\sigma}^{\dagger}c_{\br'\sigma}
+\frac{\Delta_{\br\br'}}{2}\left(c^{\dagger}_{\br\downarrow}c^{\dagger}_{\br'\uparrow}+
c^{\dagger}_{\br'\downarrow}c^{\dagger}_{\br\uparrow}\right)+h.c.
 \right],
\end{equation}
where the sum is over the nearest neighbors $\langle\br\br'\rangle$
of the (underlying) tight-binding lattice. The hopping integral
$t_{\br\br'}=t\exp(-i\bA_{\br\br'})$; in the symmetric gauge the
magnetic flux $\Phi$ enters the Peierls factor via
$\bA_{\br\br+\hat{x}}=y\Phi/2$ and $\bA_{\br\br+\hat{y}}=-x\Phi/2$,
and the $d$-wave pairing field in the vortex lattice state reads
$\Delta_{\br\br'}=\eta_{\br-\br'}\Delta_0\exp(i\theta_{\br\br'})$;
$\eta_{\bdelta}=+(-)$ if $\bdelta \parallel \hat{\bx}(\hat{\by})$.
The bond phase factors $\exp{(i\theta_{\br\br'})}$ result from a
self-consistent calculation, but the main topological feature of
$\theta_{\br\br'}$ is its $2\pi$ winding around the magnetic field
induced vortices. As the initial Ansatz for the self-consistent
solution, we choose the lattice bond variables $\theta_{\br\br'}$ as
follows: first we solve the continuum equations
$\nabla\times\nabla\phi(\br)=2\pi \hat{\bz}\sum_{i}\delta(\br-\br_i)
$ and $\nabla\cdot\nabla\phi(\br)=0$ where $\br_i$ denotes the
vortex positions which in this Letter form Abrikosov lattice. These
conditions determine $\phi(\br)$ up to $\phi_0+\bv_0\cdot \br$,
which is fixed by requiring zero overall current. Next, we define
the site variables $e^{i\phi_{\br}}$ as the value of
$\exp(i\phi(\br))$ at each site, and the initial bond variable
$e^{i\theta_{\br\br'}}$ is taken to be the geometric mean of two
neighboring site variables:
$e^{i\theta_{\br\br'}}\equiv(e^{i\phi_{\br}}+e^{i\phi_{\br'}})/|e^{i\phi_{\br}}+e^{i\phi_{\br'}}|$.
This choice guarantees that if the vortices in $e^{i\phi_{\br}}$
reside inside plaquettes, so will the vortices in
$e^{i\theta_{\br\br'}}$.

The diagonalization of the Hamiltonian ($\ref{h0}$) is equivalent to
the solution of the Bogoliubov-de~Gennes (BdG) equation
$\mathcal{\hat{H}}_0\psi_{\br}=E\psi_{\br}$ where the lattice
operator
\begin{equation}\label{bdg0}
\mathcal{\hat{H}}_0= \left(\begin{array}{cc}
\mathcal{\hat{E}}_{\br}-\mu & \hat{\Delta}_{\br}  \\
\hat{\Delta}^{\ast}_{\br} & -\mathcal{\hat{E}^{\ast}}_{\br}+\mu
\end{array}\right).
\end{equation}
Both $\mathcal{\hat{E}}_{\br}$ and $\hat{\Delta}_{\br}$ are defined
through their action on a wavefunction at the lattice site $\br$ as
$\mathcal{\hat{E}}_{\br}u_{\br}=-t\sum_{\bdelta=\pm\hat{\bx},\pm\hat{\by}}
e^{-i\bA_{\br\br+\delta}} u_{\br+\delta}$ and
$\hat{\Delta}_{\br}u_{\br}=\Delta_{0}\sum_{\bdelta} e^{i\theta_{\br
\br+\bdelta}}\eta_{\bdelta}u_{\br+\delta}$. The BdG Hamiltonian
$\mathcal{\hat{H}}_0$, as well as Pauli $\sigma$ matrices used
later, act on the two component Nambu spinor $\psi_{\br}=[u_{\br},
v_{\br}]^T$.

\underline{Translational symmetry}: While the vortex positions are
periodic, the Hamiltonian (\ref{bdg0}) is invariant only if the
discrete translations are followed by a gauge transformation
(magnetic translations). However, as shown in Refs.
\cite{FranzTesanovic2000,VafekMelikyan1_2001}, $\mathcal{\hat{H}}_0$
can be transformed into a periodic Hamiltonian by a singular gauge
transformation, $\mathcal{U}=\exp\{\frac{i}{2}\sigma_3\phi_{\br}\}$.
On a lattice, the phase factors $e^{\pm\frac{i}{2}\phi_{\br}}$ must
be defined with care. First, connect pairs of the singly quantized
vortices by branch cuts (see Fig.1). Then, choose a reference point
$\br_0$ at which $b_0$ is either one of the two solutions to
$b_0^2=e^{i\phi_{\br_0}}$, and set
$e^{\frac{i}{2}\phi_{\br_0}}=b_0$. Now, let $\br$ be a site
neighboring $\br_0$ {\em not connected by a bond that crosses a
branch cut}, and determine $e^{\frac{i}{2}\phi_{\br}}$ by the
solution to $b^2=e^{i\phi_{\br}}$ closer to
$e^{\frac{i}{2}\phi_{\br_0}}$, i.e. minimizing $|b_0-b|$. Next, set
$\br$ as a reference point and determine the phase of its neighbors
in the same way. This procedure defines $e^{\frac{i}{2}\phi_{\br}}$
on all sites of the lattice uniquely. It also ensures that the
difference of the phases of $e^{\frac{i}{2}\phi_{\br}}$ and
$e^{\frac{i}{2}\phi_{\br'}}$ lies inside
$\left(-\frac{\pi}{2},\frac{\pi}{2}\right)$ mod $2\pi$ on all bonds,
except across the bonds crossing the branch cut where it is more
than $\frac{\pi}{2}$. It follows that
\begin{equation}
e^{i\theta_{\br\br'}}e^{-\frac{i}{2}\phi_{\br}}e^{-\frac{i}{2}\phi_{\br'}}=
\frac{e^{i\phi_{\br}}+e^{i\phi_{\br'}}}{|e^{i\phi_{\br}}+e^{i\phi_{\br'}}|}
e^{-\frac{i}{2}\phi_{\br}}e^{-\frac{i}{2}\phi_{\br'}}=z_{2,\br\br'}\nonumber
\end{equation}
where $z_{2,\br\br'}=1$ on each bond except the ones crossing the
branch cut where $z_{2,\br\br'}=-1$.
\begin{figure}[t]
\includegraphics[width=0.3\textwidth]{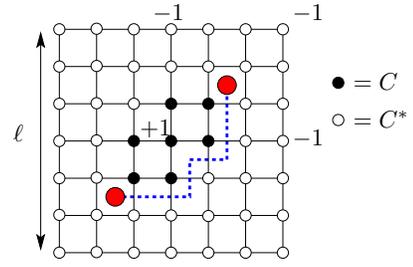}
\caption{\label{unitcell}%
Magnetic unit cell (for $\ell$=6) containing two $hc/2e$ vortices
represented by two solid (red) circles. The tight-binding lattice is
shown by the thin (black) lines, the (Z$_2$) branch-cut connecting
the vortices is shown by a dashed (blue) line. The inversion of the
branch cut about the midpoint between the vortices defines the two
regions $C$ (black dots) and $C^*$ (white dots). The gauge factor
$\gamma(\br)$ defined in the text, is 1 for the points in $C$ and
$-1$ for the points in $C^*$.}
\end{figure}
The transformed Hamiltonian,
$\mathcal{\tilde{H}}_0=\mathcal{U}^{-1}\;\mathcal{\hat{H}}_0\;\mathcal{U}$,
is now
\begin{equation}
\mathcal{\tilde{H}}_0=\sigma_3\left(
\tilde{\mathcal{E}}_{\br}-\mu\right)+\sigma_1\tilde{\Delta}_{\br},
\end{equation}
where the transformed lattice operators satisfy
\begin{eqnarray}
\mathcal{\tilde{E}}_{\br}\psi_{\br}&=&-t\sum_{\bdelta=\pm\hat{\bx},\pm\hat{\by}}
z_{2,\br\br+\delta}\times e^{i\sigma_3 V_{\br\br+\delta}} \psi_{\br+\delta}\\
\tilde{\Delta}_{\br}\psi_{\br}&=&\Delta_{0}\sum_{\bdelta=\pm\hat{\bx},\pm\hat{\by}}
z_{2,\br\br+\delta}\times \eta_{\bdelta}\psi_{\br+\delta}.
\end{eqnarray}
The physical superfluid velocity enters via the factor $ e^{i
V_{\br\br'}}
=\frac{1+e^{i(\phi_{\br'}-\phi_{\br})}}{|1+e^{i(\phi_{\br'}-\phi_{\br})}|}e^{-i\bA_{\br\br'}}$
and describes the lattice version of the semiclassical (Doppler)
effect. On the other hand, the effect of the z$_2$ field is {\em
purely quantum mechanical} and tied to the $hc/2e$ flux
quantization. It would be absent for doubly quantized $hc/e$
vortices.

The resulting Hamiltonian $\mathcal{\tilde{H}}_0$ is invariant under
discrete translations by $\ell_{x,y}$ defining the magnetic unit
cell, due to the periodicity of
$V_{\br\br'}$\cite{NielsenHedegard85} and our periodic choice of the
branch cuts. Thus, it can be diagonalized in the Bloch basis.
Extracting the crystal wavevector $\bk$ from the Bloch wavefunctions
gives $\mathcal{H}(\bk)\equiv e^{-i\bk\br}\mathcal{\tilde{H}}_0
e^{i\bk\br}$ acting on the Hilbert space of periodic Nambu spinors.

\underline{Inversion symmetry}: Our subsequent analysis will focus
on $\mathcal{H}(\bk)$. For a unit cell with $\ell_x\ell_y$ sites,
there are $2\ell_x\ell_y$ eigenvalues for each $\bk$ in the first
Brillouin zone. Consider now an inversion symmetric vortex lattice
for which $e^{i\phi_{\br}}=e^{i\phi_{-\br}}$. The inversion operator
$\mathcal{I}$, defined by its action on the wavefunctions
$\mathcal{I}\psi_{\br}=\psi_{-\br}$, transforms the Hamiltonian
according to
$\mathcal{I}\mathcal{H}(\bk)\mathcal{I}=\mathcal{H'}(-\bk)$. The
important point is that $\mathcal{H'}(-\bk)$ does not always equal
to $\mathcal{H}(-\bk)$, because the branch-cut can transform in a
non-trivial way under inversion.  A square vortex lattice depicted
in Fig. 1 represents such a case. However, the branch cut can be
restored by an additional $Z_2$ gauge transformation $\gamma_{\br}$,
where $\gamma_{\br}=1$ $\forall \br \in \mathcal{C}$, and
$\gamma_{\br}=-1$ $\forall \br \in \mathcal{C}^{\ast}$. The set
$\mathcal{C}$ is defined by the points inside the area enclosed by
the branch cut and its inverted image, and $\mathcal{C}^{\ast}$ is
its complement (see Fig. \ref{unitcell}). Thus,
$\gamma_{\br}\mathcal{I}\mathcal{H}(\bk)\mathcal{I}\gamma_{\br}=\mathcal{H}(-\bk)$.
Moreover, the spectrum of $\mathcal{H}(\bk)$ is symmetric under
$E\rightarrow -E$ for any $\mu$ \cite{VafekMelikyan2_2001}.
\begin{table}[t]
\begin{tabular}{llcccc}
\hline
\hline
 $hc/2e$
&$\bG^*$&$\br=(0,0)$&$\frac{1}{2}(\ell,0)$&$\frac{1}{2}(\ell,\ell)$&
 $\frac{1}{2}(0,\ell)$\\
\hline $\Gamma$:&$(0,0)$&1&1&-1&1\\
X:&$(-2\pi/\ell,0)$&1&-1&1&1\\
M:&$(-2\pi/\ell,-2\pi/\ell)$&1&-1&-1&-1\\
Y:&$(0,-2\pi/\ell)$&1&1&1&-1\\
\hline
\hline
$hc/e$
&$\bG^*$&$\br=(0,0)$&$\frac{1}{2}(\ell,0)$&$\frac{1}{2}(\ell,\ell)$&
 $\frac{1}{2}(0,\ell)$\\
\hline $\Gamma$:&$(0,0)$&1&-1&1&-1\\
X:&$(-2\pi/\ell,0)$&1&1&-1&-1\\
M:&$(-2\pi/\ell,-2\pi/\ell)$&1&1&1&1\\
Y:&$(0,-2\pi/\ell)$&1&-1&-1&1\\
\end{tabular}
\caption{\label{table} The value of the diagonal elements of
$\mathcal{P}_{\bk^*}=(-1)^{x+y}e^{i\bG^*\cdot\br}\gamma(\br)\mathcal{I}\mathbbm{1}$
at invariant points for the case in Fig.\ref{unitcell}. For $hc/2e$
vortices, $\gamma_{\br}$ is non-trivial and
Tr$\mathcal{P}_{\bk^*}=\pm4$ for all $\bk^*$ indicated in the
leftmost column. This guarantees at least 16 zeros in the Brillouin
zone. On the other hand, for $hc/e$ vortices $\gamma_{\br}=1$ and
Tr$\mathcal{P}=8$ only at $\bk^*=M$, which guarantees only 8 zeros.
For band notation see Fig.{\ref{bands}}. }
\end{table}

\underline{Index theorem:} For $\mu=0$ we can define
\begin{equation}
\mathcal{P}=(-1)^{x+y}\gamma_{\br}\mathcal{I}\; \mathbbm{1},
\label{P}
\end{equation}
where $\mathbbm{1}$ is the unity in the Nambu space and the function
$(-1)^{x+y}$ changes sign on every other site. Clearly,
$\mathcal{P}^2=1$. In addition this operator satisfies
\begin{equation}
\mathcal{P}\mathcal{H}(\bk)\mathcal{P}=-\mathcal{H}(-\bk),
\end{equation}
and, in particular, at $\bk=0$,
$\mathcal{P}\mathcal{H}(0)\mathcal{P}=-\mathcal{H}(0)$. Therefore,
if there are zero modes at $\bk=0$, they can be chosen to be
eigenstates of $\mathcal{P}$ with eigenvalues $\pm1$. At $\bk=0$,
for any eigenstate $\psi_{E,\br}$ of $\mathcal{H}(0)$ with energy
$E\neq 0$, we can generate an eigenstate
$\psi_{-E,\br}=\mathcal{P}\psi_{E,\br}$ with energy $-E$. Therefore,
the only diagonal matrix elements of $\mathcal{P}$ in the basis of
the eigenstates of $\mathcal{H}(0)$ come from the zero-energy
states. Let us denote by $n_{\pm}$ the number of zero-energy states
with $\mathcal{P}$-eigenvalue $\pm1$. Then,
$\mbox{Tr}{\mathcal{P}}=n_{+}-n_-$. Therefore,
$\mbox{Tr}{\mathcal{P}}$ constitutes the lower bound on the number
of zero energy states of $\mathcal{H}_0$.

Since $\mbox{Tr}{\mathcal{P}}$ is independent of the basis, we can
compute the trace in the coordinate basis. Consider again the square
vortex lattice depicted in Fig. 1 with two $hc/2e$ vortices per unit
cell with the primitive vortex cell 45$^o$ relative to the
tight-binding lattice and the magnetic length $\ell=2(2n+1)$. Then,
there are only four spatial points which remain invariant under
inversion and contribute to the trace. We listed the values of
$\mathcal{P}$ in Table \ref{table}, from which it follows that
$\mbox{Tr}\mathcal{P}=4$. Therefore, the index of $\mathcal{H}(0)$
is 4 and it has at least 4 zero-energy eigenstates. Same argument
holds for $\mathcal{H}(\bk^{\ast})$ where $\bk^{\ast}\in \{(0,0),
(\pi/\ell,0),(\pi/\ell,\pi/\ell),(0,\pi/\ell)\}$, with the only
modification being
$\mathcal{P}_{\bk^*}=(-1)^{x+y}e^{i\bG^*\cdot\br}\gamma_{\br}\mathcal{I}1_2$,
where $\bG^*=-2\bk^*$. {\em So for $l=2(2n+1)$ there are at least 16
zero energy eigenstates in the first Brillouin zone}.

Consider now the same vortex arrangement, but for $hc/e$ vortex
lattice (or for $hc/2e$ but ignoring the $Z_2$ branch cuts). Then,
$\gamma_{\br}=1$ everywhere and $\mbox{Tr}\mathcal{P}_{\bk^*}=0$
except at M point ($\bk^*=(\pi/\ell,\pi/\ell)$) where
$\mbox{Tr}\mathcal{P}_{\bk^*}=8$, i.e. the index theorem guarantees
only 8 zero energy states. This is the same number of zeros as in
the absence of magnetic field. Thus, while the effects of the
superflow are perturbative, the effect of the branch cuts is
non-perturbative. The symmetries discussed exist only if the bonds
change sign exactly. Clearly, 16 zero energy states cannot be
reached by perturbation theory around a state with 8 zero energy
states, such as the $d$-wave superconductor in the absence of the
magnetic field.

On the other hand, for $\ell=2(2n)$, $\mbox{Tr}\mathcal{P}=0$ in
both cases, because there are no lattice points invariant under
inversion. The index theorem thus does not guarantee any zero energy
states, i.e. {\em the spectrum is gapped.}

\begin{figure}[t]
\includegraphics[width=0.4\textwidth,clip]{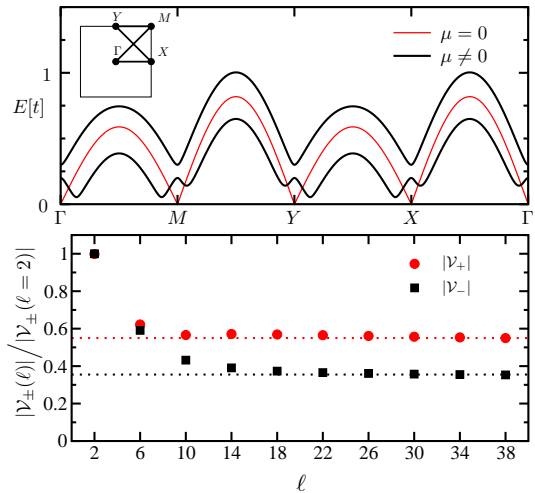}
\caption{Upper panel: The energy dispersion in the gapless (nodal)
($\mu=0$) and gapped ($\mu\neq 0$) phases. Only the positive
energies are shown as the spectrum is symmetric under $E\rightarrow
-E$. For $\mu=0$ and $\ell=4n+2$ each (red) band is doubly
degenerate and there are 16 zero modes in the Brillouin zone. Finite
$\mu$ splits the degeneracy and the low energy spectrum corresponds
to 8 massive Dirac fermions, each contributing equally to the
spin Hall conductivity $\sigma^{s}_{xy}=\pm4\frac{\hbar}{8\pi}$.\\
Lower panel: The renormalized nodal velocities normalized to their
value at $\ell=2$ as functions of $\ell=4n+2$, approach the
universal value dependent on $\alpha=\Delta_0/t$ only. Here
$\alpha=0.1$. \label{bands}}
\end{figure}

\underline{Degenerate Dirac cones}: Around each of the zeros in the
magnetic Brillouin zone, the energy vanishes linearly with the
deviation $\delta \bk$ from the degeneracy points. This follows from
the standard $\bk\cdot \bp$ perturbation theory arguments as well as
the fact that at $\mu=0$ all the bands are doubly degenerate
\cite{VafekMelikyan2_2001}. Expanding the Hamiltonian
$\mathcal{H}(\delta \bk)$ in the basis of the zero energy states
near $\bk=0$ we find
\begin{equation}
\label{Heff} \mathcal{H}_{eff}=\mathcal{V}_+\delta k_+\Gamma_1 +
\mathcal{V}_-\delta k_-\Gamma_2,
\end{equation}
where $\Gamma_{1,2}$ are 4$\times$4 Dirac gamma matrices and $\delta
k_{\pm}=(k_x\pm k_y)/\sqrt{2}$. The spectrum near $\bk^*=0$ thus
consists of two doubly degenerate Dirac cones with energies
\begin{equation}
\mathcal{E}_0=\pm t\sqrt{\mathcal{V}_+^2\delta
k_+^2+\mathcal{V}_-^2\delta k_-^2}. \label{disp}
\end{equation}
The new nodal velocities $\mathcal{V}_{\pm}$ must be determined from
the wavefunctions of the degenerate multiplet. For the square vortex
lattice in Fig.\ref{unitcell}, we found analytically that for
$\ell=2$,
$\mathcal{V}_{\pm}^2=2(1+\alpha^2)\pm\frac{2\alpha^2}{1+\alpha^2}(\alpha^2-2(1+\sqrt{2}))$,
where the bare Dirac cone anisotropy $\alpha=\Delta_0/t$.
Eq.(\ref{disp}) holds also near $\bk^*=(\pi/\ell,\pi/\ell)$, while
near $(\pi/\ell,0)$ and $(0,\pi/\ell)$, $\delta k_{\pm} \rightarrow
\pm\delta k_{\mp}$ as required by the fourfold symmetry. While
$\ell=2$ corresponds to an unrealistically large magnetic field for
physical parameters, $\ell=2$ is well representative of the sequence
$\ell=4n+2$ and has the virtue of being analytically soluble. We
verified numerically that the qualitative results discussed below
are indeed the same along the sequence $\ell=4n+2$, for integer $n$.
It is not difficult to see that the effective Hamiltonian
(\ref{Heff}) has the same form for any $\ell=4n+2$. In addition, in
the limit of large $n$, the Dirac velocities $\mathcal{V}_{\pm}$
become universal functions of $\alpha$ in accordance with Simon-Lee
scaling \cite{SimonLee1997}. We verified this numerically (see
Fig.\ref{bands}).

\underline{P-h symmetry breaking and topology of the gaps}: For
small but finite $\mu$, all the degeneracies are lifted
Fig.(\ref{bands}), but the spectrum is still symmetric under
$E\rightarrow -E$ for each $\bk$ point due to the inversion symmetry
of the vortex lattice \cite{VafekMelikyan2_2001}. At low energies we
have
\begin{equation}
\mathcal{E}=\pm\sqrt{\mathcal{E}^2_0+m^2_0\pm
|\mu|\sqrt{\tilde{\mathcal{E}}^2+\tilde{m}^2_0}},
\end{equation}
where the $\pm$ signs are uncorrelated, $m_0$ and $\tilde{m}_0$ are
functions of $\ell$ and $\alpha$, and
$\tilde{\mathcal{E}}^2=t^2\left(\tilde{\mathcal{V}}_+^2\delta
k_+^2+\tilde{\mathcal{V}}_-^2\delta k_-^2 \right)$. As shown in
Fig.(\ref{bands}), at each $\bk^*$ there is a local {\em maximum} in
the lowest positive band of the spectrum; the local minimum is split
and moved to $\{\delta k_+,\delta k_-\}=\{\pm k_{min},0\}$ near
$(0,0)$ and $(\pi/\ell,\pi/\ell)$, and $\{\delta k_+,\delta
k_-\}=\{0,\pm k_{min}\}$ near $(\pi/\ell,0)$ and $(0,\pi/\ell)$,
where
$k_{min}=\frac{\mu}{2t}\frac{1}{\tilde{\mathcal{V}}_+}\sqrt{\frac{\tilde{\mathcal{V}}_+^4}{\mathcal{V}_+^4}-
\frac{4\tilde{m}^2_0}{\mu^2}}$.

In the vicinity of each of the 8 band-gap minima, the spectrum is
equivalent to the 2+1 dimensional massive Dirac particle, with mass
$m_D$ given by the gap minimum $E_{min}$. To see that, around each
of the eight local energy minima, project onto the lowest energy
basis, $|\pm\rangle$, such that
$\mathcal{H}(\bk_{min})|\pm\rangle=\pm E_{min}|\pm\rangle$, and
expand the resulting effective Hamiltonian in $\delta \bk$:
\begin{equation}
\tilde{\mathcal{H}}_{eff}=E_{\min}\tau_3+v_+\delta
k_+\tau_1+v_-\delta k_-\tau_2, \label{hmassive}
\end{equation}
$\tau_{i}$ are the Pauli matrices acting on the 2-dimensional
subspace spanned by $|\pm\rangle$. Clearly, $\tilde{\mathcal{H}}$ is
equivalent to the Hamiltonian of a 2+1 dimensional {\em massive}
Dirac particle. By the $E\rightarrow -E$ symmetry of the spectrum,
the negative energy states are all occupied. It is well known, that
in 2+1 dimensions, in the representation where $v_{\pm}>0$, the sign
of the mass determines the Hall conductivity
$\sigma_{xy}=-\frac{1}{2}\mbox{sgn}(m_D)\frac{e^2}{h}$
\cite{Haldane1988,Tesanovic1989,Ludwig1994}. In the case at hand,
the charge is not conserved due to broken U(1) symmetry of the
superconductor, but the component of the spin along the magnetic
field is conserved, and so spin Hall conductivity,
$\sigma^{s}_{xy}$, is quantized
\cite{Volovik89,Senthil1999,Ashvin2001,VafekMelikyan2_2001}. We
verified that all 8 of the Dirac particles have the same chirality,
i.e. their additive contributions are the same. In order to
determine the {\em total} value of $\sigma_{xy}$, generically, it is
not enough to know the low energy part of the spectrum. Only the
change of $\sigma^{s}_{xy}$ can thus be determined. Nevertheless, in
this case, at $\mu=0$, $\sigma_{xy}=0$ due to the particle-hole
symmetry \cite{VafekMelikyan2_2001} and since the change is $\pm 8$,
$\sigma^{s}_{xy}=\pm4$ (in appropriate units). We verified this
numerically as well. The resulting state at $\mu\neq0$ remains
distinguishable from the case of $hc/e$ vortex lattice, where
$|\sigma_{xy}|\leq 2$.

While the above analysis pertains to the initial Ansatz for
$e^{i\theta_{\br\br'}}$, the main conclusions depend only on the
translational and inversion invariance of $\Delta_{\br\br'}\times
z_{2,\br\br'}e^{-\frac{i}{2}\phi_{\br}}e^{-\frac{i}{2}\phi_{\br'}}$
which we verified holds at each step of the self-consistent
iteration for a model with the nearest neighbor attraction that
stabilizes d-wave pairing.

Finally, we conclude with comments about experiments. Recently it
was observed that the low temperature thermal conductivity,
$\kappa_{xx}$, of the ultraclean YBa$_2$Cu$_3$O$_7$ showed
pronounced deviations from the semiclassical predictions
\cite{Hill2004}. Rather, it was qualitatively consistent with the
onset of a new universal $\kappa_{xx}$, distinct from its zero field
value. As observed by Franz and one of us (O.V), the universal
renormalization of the nodal velocities, shown in Fig.\ref{bands},
could provide a qualitative explanation \cite{Franz2001}. Presently,
measurements of $\kappa_{xy}$ extend to about 10K which is too high
to observe the quantization discussed.

We acknowledge helpful discussions with Professor Z.
Te\v{s}anovi\'{c}. O.V. would like to thank the Aspen Center for
Physics for their hospitality where part of this work was completed.
O.V. was supported by the Stanford ITP fellowship.
\bibliography{bibliography}
\end{document}